\documentclass[useAMS,usenatbib,usegraphicx]{mn2e}
\bibpunct{(}{)}{;}{a}{}{,}
\usepackage{txfonts,longtable}
\title[Molecular Gas in Early-Type Galaxies] {Molecular Gas in SAURON
    Early-Type Galaxies: Detection of $^{13}$CO and HCN
    Emission\thanks{Based on observations carried out with the IRAM
    30m telescope. IRAM is supported by INSU/CNRS (France), MPG
    (Germany) and IGN (Spain).}}
  \author[M.\ Krips et al.]
  {M.\ Krips,$^1$\thanks{E-mail: krips@iram.fr} A.\ F.\ Crocker,$^2$ M.\
    Bureau,$^2$ F.\ Combes,$^3$
    and L.\ M.\ Young$^4$\\
    $^1$Institut de Radio Astronomie Millimetrique (IRAM), Domaine
    Universitaire, 300 rue de la Piscine, 38406 Saint Martin d'H\`eres, France\\
    $^2$Sub-Department of Astrophysics, University of Oxford, Denys
    Wilkinson Building, Keble Road, Oxford OX1 3RH, U.K.\\
    $^3$Observatoire de Paris, LERMA, 61 Av.\ de l'Observatoire, 75014 Paris, France\\
    $^4$Physics Department, New Mexico Institute of Mining and
    Technology, Socorro, NM 87801, U.S.A.
    }
\date{Received ; accepted}
\begin{document}
\maketitle
%
%
\begin{abstract} 

In a pilot project to study the relationship between star formation
and molecular gas properties in nearby normal early-type galaxies, we
have obtained observations of dense molecular gas tracers in the four
galaxies of the {\tt SAURON} sample with the strongest $^{12}$CO
emission.  We used the Institut de Radio Astronomie Millimetrique
(IRAM) 30m telescope $3$ and $1$~mm heterodyne receivers to observe
$^{13}$CO(J=$1$--$0$), $^{13}$CO(J=$2$--$1$), HCN(J=$1$--$0$) and
HCO$^+$(J=$1$--$0$). We report the detection of $^{13}$CO emission in
all four {\tt SAURON} sources and HCN emission in three sources, while
no HCO$^+$ emission was found to our detection limits in any of the
four galaxies.  We find that the $^{13}$CO/$^{12}$CO ratios of three
{\tt SAURON} galaxies are somewhat higher than those in galaxies of
different Hubble types. The HCN/$^{12}$CO and HCN/$^{13}$CO ratios of
all four {\tt SAURON} galaxies resemble those of nearby Seyfert and
dwarf galaxies with normal star formation rates, rather than those of
starburst galaxies. The HCN/HCO$^+$ ratio is found to be relatively
high (i.e., $>$1) in the three {\tt SAURON} galaxies with detected HCN
emission, mimicking the behaviour in other star-forming galaxies but
being higher than in starburst galaxies. When compared to most
galaxies, it thus appears that $^{13}$CO is enhanced (relative to
$^{12}$CO) in three out of four {\tt SAURON} galaxies and HCO$^+$ is
weak (relative to HCN) in three out of three galaxies.

All three galaxies detected in HCN follow the standard HCN--infrared
luminosity and dense gas fraction--star formation efficiency
correlations. As already suggested by $^{12}$CO observations, when
traced by infrared radiation, star formation in the three {\tt SAURON}
galaxies thus appears to follow the same physical laws as in galaxies
of different Hubble types. The star formation rate and fraction of
dense molecular gas however do not reach the high values found in
nearby starburst galaxies, but rather resemble those of nearby normal
star-forming galaxies.
\end{abstract}
\begin{keywords}
galaxies: elliptical and lenticular, cD~-- galaxies: evolution~--
galaxies: ISM~-- galaxies: kinematics and dynamics~-- galaxies:
structure~-- Radio lines: galaxies
\end{keywords}

\section{INTRODUCTION}
\label{sec:intro}
%
%
%
%

\begin{figure*}
\begin{center}
\resizebox{\hsize}{!}{\rotatebox{-90}{\includegraphics{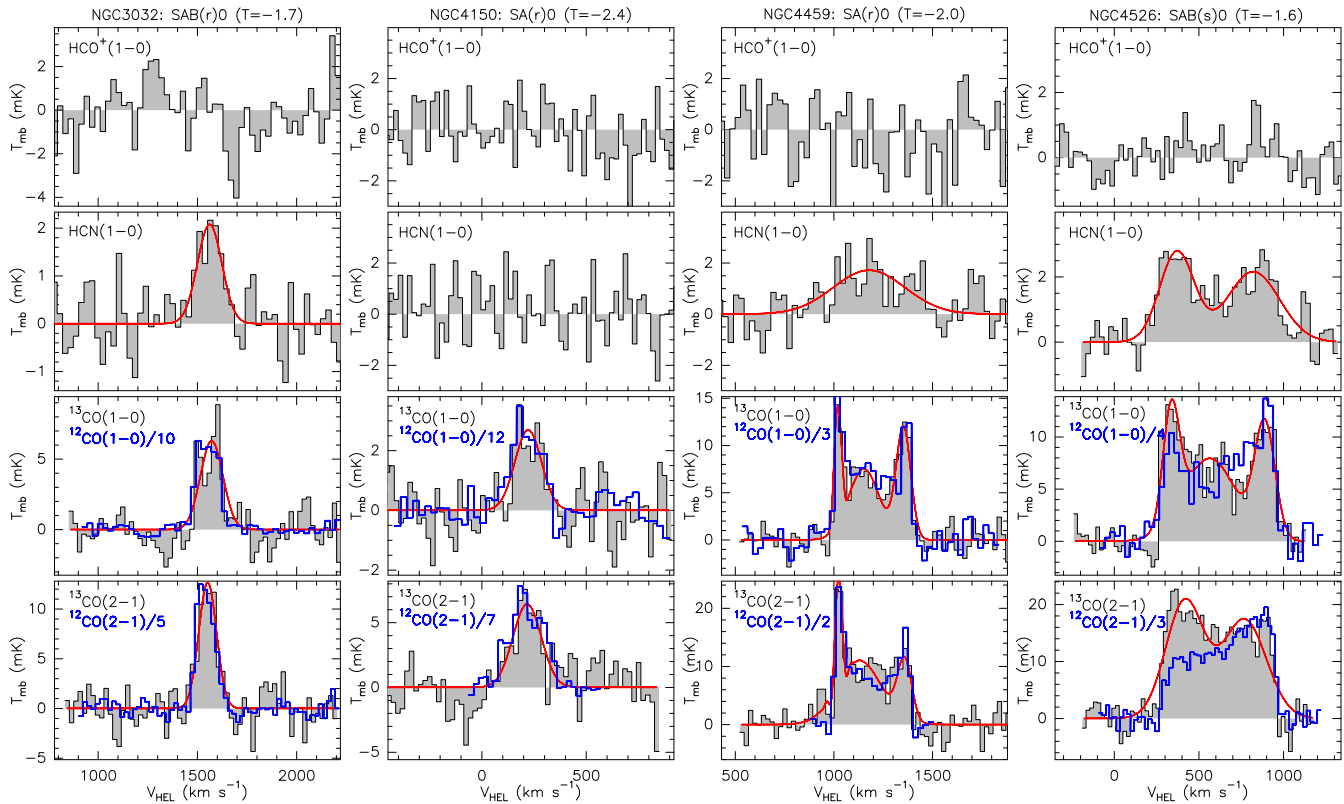}}}
\caption{Emission line spectra of the four {\tt SAURON} early-type
  galaxies observed. From top to bottom in each panel:
  HCO$^+$(J=1--0), HCN(J=1--0), $^{13}$CO(J=1--0) and
  $^{13}$CO(J=2--1).  From left to right: NGC~3032, NGC~4150, NGC~4459
  and NGC~4526.  The $^{13}$CO emission is overlaid with the spectra
  of the $^{12}$CO emission ({\it blue}) from \citet{com07}, scaled
  down to match the $^{13}$CO line emission; the scaling factors are
  indicated in the respective spectra. The (single, dual, triple)
  Gaussian fits to the line spectra are plotted in red.}
\label{fig:fig1}
\end{center}
\end{figure*}

Over the last twenty years, growing evidence has accumulated that
nearby early-type galaxies (E/S0s) are not always devoid of molecular
gas, and can even harbour a substantial amount of it
\citep*[e.g.,][]{lee91,wik95,bet03,sag07,com07,you09}, often settled
in a regularly rotating disc
\citep*[e.g.,][]{you02,you05,you08,cro08,cro09,cro10}. Interestingly,
not all early-type galaxies with a substantial amount of molecular gas
show obvious signs of (current) star formation \citep[SF;
  e.g.,][]{cro08,cro09,cro10,you08}.  However, the converse is
generally true \citep[e.g.,][]{jeo09}, and the $\ga30$~per cent
fraction of SF objects within the nearby early-type galaxy population
\citep[e.g.,][]{yi05} appears to be superficially consistent with
their $\approx25$~per cent CO detection rate
\citep[e.g.,][]{sag06,sag07,com07,you09}. This fact raises many as yet
unanswered questions: how is the (current) SF ignited in early-type
galaxies? Is SF coupled to the molecular discs, i.e.\ do all molecular
gas discs form stars? Does SF only take place in the densest regions,
perhaps ill-traced by the $^{12}$CO lines used in most surveys?  How,
if at all, does the morphology of the molecular discs and the physical
characteristics of the molecular gas correlate with the ages of the
stars? In order to answer some of these questions, it is essential to
study the spatially- and kinematically-resolved properties of the
stellar populations, as well as those of the molecular gas in
early-type galaxies (E/S0).  CO emission can also easily trace signs
of gravitational interactions or other dynamical perturbations that
can help to understand how star formation is induced.

Observations of a sample of 48 nearby E/S0 galaxies have been carried
out with a custom-designed panoramic optical integral-field
spectrograph \citep{bac01} within the framework of the {\tt SAURON}
project. A description of the sample and illustrative results can be
found in \citet{zee02}. Results include the detection of kinematic
misalignments, twists and decoupled cores as well as central discs in
roughly half of the sources \citep{ems04,kra08}. Information on the
ionized gas (distribution, kinematics and ionisation;
\citealt{sar06,sar10}) and the age, metallicity and alpha-element
enhancement of the stellar populations \citep[][]{kun06,kun10} are
also available. The specific angular momentum of the galaxies appears
to be a dominant factor in their evolution \citep{ems07,cap07}, and
all CO detections are found in fast-rotating systems \citep{you09}.

In order to study star formation in the {\tt SAURON} sources and
understand the various phenomena observed in the stellar and ionised
gas data, observations looking for central molecular gas in the nearby
{\tt SAURON} E/S0 galaxies were obtained by \citet{com07}.  $^{12}$CO
emission was detected in $12$ of $43$ galaxies observed with the
Institut de Radio Astronomie Millimetrique (IRAM) 30m telescope,
suggesting typical gas masses in a range between a few 10$^{6}$ to
10$^{8}$ M$_\odot$.  \citet{com07} further found that the amount of
molecular gas in the galaxies correlates with their far-infrared (FIR)
luminosities, an often-used indicator of star formation. The gas-rich
sources show the most pronounced star formation, as for normal
star-forming spiral galaxies. However, the lack of correlations
between the molecular gas and most stellar properties suggests that
the molecular gas is largely unrelated to the old, pre-existing
stellar population, and might have been externally accreted, as a
result for instance of a merger or interaction event \citep{com07}.

%
%
\begin{table}
\caption{Basic parameters of the four {\tt SAURON}
  galaxies.}
\label{tab:tab1}      
\centering
\begin{tabular}{lrrrrc}     
\hline
NGC  & RA(J2000)      & DEC(J2000) & $V_\odot$    & D       & Activity \\
         & (hh:mm:ss.s) & (dd:mm:ss) & (km~s$^{-1}$) & (Mpc) & Type     \\
\hline 
3032 & 09:52:08.2 & $+$29:14:10 & 1533 & 21.7 & H~{\small II}         \\
4150 & 12:10:33.6 & $+$30:24:06 &  226 & 13.7 & none        \\
4459 & 12:29:00.0 & $+$13:58:43 & 1210 & 16.3 & H~{\small II}+L   \\
4526 & 12:34:03.0 & $+$07:41:57 &  550 & 16.3 & none        \\
\hline
\end{tabular}
 
Notes: Coordinates, distances, velocities and activity types are taken
  from the NASA/IPAC Extragalactic Database (NED). Galaxy types:
  L=LINER, H~{\small II}=star formation bursts.
\end{table}

%
%
\begin{table*}
\caption{Properties of the molecular line spectra.}
\label{tab:tab2}      
\centering          
\begin{tabular}{lrrrrrr}     
\hline
\\[-0.2cm]                
Line   & $^{12}$CO(J=1--0)  & $^{12}$CO(J=2--1) 
   & $^{13}$CO(J=1--0)  & $^{13}$CO(J=2--1) 
   & HCN(J=1--0) & HCO$^+$(J=1--0)$^b$ \\
Parameters$^a$   & (CYB07) & (CYB07) & (this paper) & (this paper) & (this paper) & (this paper)\\[0.2cm]
\hline    
\\[-0.2cm]                
NGC3032: & \\  
\hskip 0.4cm $I$ (K~km~s$^{-1}$)         & 9.3$\pm$0.3 & 7.9$\pm$0.2  & 0.9$\pm$0.1 & 1.4$\pm$0.2 & 0.32$\pm$0.09 & $<$0.2\\
\hskip 0.4cm $<v_\odot>$ (km~s$^{-1}$)   & 1549$\pm$2  & 1537$\pm$2   & 1571$\pm$10 & 1561$\pm$10 & 1571$\pm$13   &  \\
\hskip 0.4cm $\Delta v$ (km~s$^{-1}$)    & 129$\pm$4   & 103$\pm$3    &  130$\pm$14 &  104$\pm$10 &  160$\pm$30   &  \\
\\[-0.2cm]                
NGC4150: & \\
\hskip 0.4cm $I$ (K~km~s$^{-1}$)         & 6.1$\pm$0.5 & 13.2$\pm$0.5 & 0.4$\pm$0.1 & 1.1$\pm$0.2 & $<$0.2  & $<$0.2 \\
\hskip 0.4cm $<v_\odot>$ (km~s$^{-1}$)   & 204$\pm$7   & 200$\pm$3    & 221$\pm$14  & 226$\pm$12  &               &  \\
\hskip 0.4cm $\Delta v$ (km~s$^{-1}$)    & 158$\pm$14  & 189$\pm$8    & 150$\pm$30  & 160$\pm$30  &               &  \\
\\[-0.2cm]                
NGC4459: & \\
\hskip 0.4cm $I$ (K~km~s$^{-1}$)         & 10.9$\pm$0.5& 14.3$\pm$0.5 & 3.2$\pm$0.2   & 4.9$\pm$0.4 & 0.8$\pm$0.1 & $<$0.7\\
\hskip 0.4cm $<v_\odot>$ (km~s$^{-1}$)   & 1169$\pm$9  & 1176$\pm$8   & 1194$\pm$22   & 1179$\pm$40 & 1140$\pm$20 &  \\
\hskip 0.4cm $\Delta v$ (km~s$^{-1}$)    & 477$\pm$16  & 405$\pm$15   & 390$\pm$20    & 380$\pm$20  & 440$\pm$70  &  \\
\\[-0.2cm]                
NGC4526: & \\
\hskip 0.4cm $I$ (K~km~s$^{-1}$)         & 23.8$\pm$0.8& 37.4$\pm$0.7 & 6.0$\pm$0.2   & 11$\pm$0.4 & 1.6$\pm$0.1 & $<$0.7\\
\hskip 0.4cm $<v_\odot>$ (km~s$^{-1}$)   & 697$\pm$2   & 695$\pm$5    & 605$\pm$30    & 600$\pm$30  & 606$\pm$40  & \\
\hskip 0.4cm $\Delta v$ (km~s$^{-1}$)    & 650$\pm$23  & 533$\pm$11   & 687$\pm$26    & 677$\pm$27  & 760$\pm$50  & \\
\\[-0.2cm]                
\hline                  
\end{tabular}

Notes: $^a$Line properties have been obtained by fitting a single
Gaussian to the data from \citet{com07} and this paper;
$I\equiv\int~T_{\rm mb}~dv$ is the velocity-integrated line intensity;
$<v_\odot>$ is the mean (central) heliocentric velocity; $\Delta v$ is
the velocity width (FWHM). $^b$ Determined from 3$\sigma$ upper limits
from spectrum at $\sim$100~km~s$^{-1}$ spectral resolution assuming
the same FWHM as found for the HCN(J=1--0) or $^{13}$CO(J=1--0) line.
\end{table*}

However, as $^{12}$CO emission has been found to be a rather
unreliable tracer of the dense gas where star formation takes place
\citep[e.g.,][]{gao04a,gao04b}, as a pilot study we focus here on the
volume and/or column density tracers HCN, HCO$^+$ and $^{13}$CO in the
four {\tt SAURON} galaxies with the brightest $^{12}$CO
emission. Basic parameters of these four sources are given in
Table~\ref{tab:tab1}.  They form an interesting subset of the {\tt
SAURON} sample as they nicely populate the two main evolutionary
groups: i) those with a more 'troubled' recent past, likely involving
accretion and/or minor interactions with other galaxies (NGC~3032 and
NGC~4150); and ii) those with a more peaceful immediate past dominated
by secular evolution (NGC~4459 and NGC~4526). The first group appears
to harbor a large fraction of widely distributed young stars while the
second group possesses a lower young stellar fraction that is
spatially limited to the center \citep[][]{kun10}. Additionally,
NGC~3032 shows signs that its young stars have a lower metallicity
than its bulk stellar population. Nevertheless, only NGC~3032 and
NGC~4459 show signs of circumnuclear activity both in the form of
H~{\small II} regions, indicative of star formation bursts, and, in
the case of NGC~4459 also in the form of a low-ionization nuclear
emission region (LINER) nucleus.

The paper is organised as follows. Section~\ref{sec:obs} presents the
observations. Section~\ref{sec:results} consists of the results and a
discussion of the data and their trends. Finally,
section~\ref{sec:conclusions} summarizes our findings and conclusions.

%
%
\begin{table*}
\caption{Molecular line ratios for a representative set of nearby
 galaxies covering the Hubble sequence and different activity types.}
\label{tab:tab3}      
\begin{center}
\begin{tabular}{lcccccccc}     
\hline       
Source & Activity Type$^a$ &  $\frac{^{13}{\rm CO(J=1-0)}}{^{12}{\rm CO(J=1-0)}}$  
            & $\frac{^{13}{\rm CO(J=2-1)}}{^{12}{\rm CO(J=2-1)}}$  
            & $\frac{{\rm HCN(J=1-0)}}{^{13}{\rm CO(J=1-0)}}$  
            & $\frac{{\rm HCN(J=1-0)}}{^{12}{\rm CO(J=1-0)}}$  
            & $\frac{{\rm HCN(J=1-0)}}{{\rm HCO^+(J=1-0)}}$  
          & Ref$^b$  & Region$^c$ \\
\hline    
\multicolumn{9}{c}{Ellipticals: $-$6 $\leq$ T  $\leq$ $-$4}\\[0.1cm]
Cen~A    &   Sy2 & $\sim$0.07 &  $\sim$0.08  & $\sim$0.9 & $\sim$0.06 & $\sim$0.6$^d$ & (1) & Centre \\
        & & $\sim$0.1  &  $\sim$0.09  & $\sim$0.2 & $\sim$0.02 &  --           & (1) & Dust Lane \\[0.2cm]
\multicolumn{9}{c}{Sauron (S0): $-$4 $<$ T $\leq$ 0}\\[0.1cm]

NGC~3032 &   H~{\small II}   & 0.09$\pm$0.01 & 0.20$\pm$0.02 & 0.36$\pm$0.07 & 0.03$\pm$0.01 & $>$1.7 & (2) & Centre \\  
NGC~4150 &   NONE  &  0.07$\pm$0.01 & 0.08$\pm$0.01 &  $<$0.5   &  $<$0.03          & --     & (2) & Centre \\
NGC~4459 &   H~{\small II}+L &  0.30$\pm$0.02 & 0.34$\pm$0.02 & 0.23$\pm$0.03 & 0.07$\pm$0.01 & $>$1.1 & (2) & Centre \\
NGC~4526 &   NONE  &  0.26$\pm$0.01 & $\sim$0.3        & 0.35$\pm$0.02 & 0.09$\pm$0.01 & $>$2.3 & (2) & Centre \\[0.15cm]
\multicolumn{9}{c}{Seyferts: 0 $<$ T $<$ +8}\\[0.1cm]

NGC~1068$^\star$  &   Sy2        & 0.09$\pm$0.02   & 0.06$\pm$0.01   & 1.6$\pm$0.6    & 0.13$\pm$0.02   & 1.7$\pm$0.2  & (3,6) & Centre\\
                  &            & $\sim$0.02      & 0.062$\pm$0.002 & $\sim$80       &  $\sim$1.4      & $\sim$1.4    & (4,19) & CND$^d$\\
NGC~2237          &   ?          & $\sim$0.1       & --              & 0.41$\pm$0.07  & 0.04$\pm$0.01   & --          & (4,5)   & Centre \\
NGC~3079          &   Sy2/L      &   0.06$\pm$0.01   &   0.07$\pm$0.01   &   0.6$\pm$0.2    &   0.039$\pm$0.09  &   $>$5        &   (6,23)  &   Centre \\
NGC~3627          &   Sy2/L      & $\sim$0.06      & --              & $\sim$1.1      & $\sim$0.07      & 1.0$\pm$0.1 & (3,8)   & Centre \\
NGC~3628          &   L+H~{\small II}      &   0.12$\pm$0.02   &   0.10$\pm$0.02   &   0.34$\pm$0.09  &   0.027$\pm$0.007 &   1.6$\pm$0.5 &   (22,26) &   Centre \\
NGC~3982          &   Sy1.9+H~{\small II}  & 0.07$\pm$0.01   & --              & 0.32$\pm$0.08  & 0.022$\pm$0.005 & --          & (4)     & Centre \\
NGC~4051          &   Sy1/NL     & 0.06$\pm$0.01   & --              & 0.32$\pm$0.08  & 0.019$\pm$0.004 & --          & (4)     & Centre \\
NGC~4258          &   Sy1.9/L    & 0.10$\pm$0.01   & --              & 0.29$\pm$0.03  & 0.029$\pm$0.003 & --          & (4)     & Centre \\
NGC~4388          &   Sy2        & 0.03$\pm$0.01   & --              & 0.3$\pm$0.1    & 0.010$\pm$0.003 & --          & (4)     & Centre \\
NGC~4826          &   Sy2+H~{\small II}    & 0.12$\pm$0.02   & 0.13$\pm$0.02   & 0.40$\pm$0.04  & 0.06$\pm$0.01   & 1.7$\pm$0.1 & (3,6)   & Centre\\
NGC~4945          &   Sy2        &   0.06$\pm$0.01   &   0.11$\pm$0.01   &   0.75$\pm$0.02  &   0.05$\pm$0.01   &    $\sim$1 &   (27,28) &   Centre \\
NGC~5033$^\star$  &   Sy1.8      & 0.12$\pm$0.01   & --              & 0.36$\pm$0.03  & 0.044$\pm$0.005 & $\sim$1.9   & (4)     & CND$^d$ \\
NGC~5194$^\star$  &   Sy2        & $\sim$0.1-0.2   & $\sim$0.1       & $\sim$2        & $\sim$0.5       & 1.4$\pm$0.1 & (3,7,8) & CND$^d$ \\
NGC~6951$^\star$  &   Sy2/L      &  --             & 0.13$\pm$0.02   & $\gtrsim$2     & $\gtrsim$2      & 1.4$\pm$0.1 & (3,5)   & CND$^d$ \\
NGC~7172          &   Sy2+H~{\small II}    & $\sim$0.1-0.2   & --              & 0.33$\pm$0.07  & $\sim$0.03      & --          & (4)     & Centre \\
NGC~7314          &   Sy1.9      & $\sim$0.1-0.2   & --              & $<$0.4         & $\sim$0.05      & --          & (4)     & Centre \\
NGC~7331          &   L          &   $\sim$0.14      &   $\sim$0.17      &   $\sim$0.03     &   $\sim$0.04      &   $>$1.7        &   (22,24) &   Center \\
NGC~7582          &   Sy2        &   --              &   --              &   --             &   0.03$\pm$0.01   &   1.3$\pm$0.6   &   (29)    &   Centre \\
NGC~7469$^\star$     &   Sy1.2   &   $\sim$0.06  &   $\sim$0.05      &   0.95$\pm$0.1   &   $\sim$0.1       &   $\sim$1.4     &   (4,6,7) &   Centre \\
Mrk~231$^{\dag\dag}$ &   Sy1     &   --          &   $<$0.03         &   --             &   $\sim$0.2       &   0.72$\pm$0.09 &   (29,30) &   Centre \\
Mrk~331$^\dag$       &   Sy2+H~{\small II} &   --          &   --              &   --             &   $\sim$0.2       &   $\sim$1.3     &   (29)    &   Centre \\
IRAS~05414+5840$^\dag$ &   Sy2   &   --          &   --              &   --             &   0.040$\pm$0.006 &   0.6$\pm$0.2   &   (29)    &   Centre\\[0.15cm]

\multicolumn{9}{c}{Starbursts: 0 $<$ T $<$ +8}\\[0.1cm]
M82               &   SB+H~{\small II}       & $\sim$0.03-0.1 & $\sim$0.1           & $\sim$1       & $\sim$0.2      & 0.7$\pm$0.1   & (3,8,9)   & Centre \\
M83               &   SB+H~{\small II}       &   0.10$\pm$0.01  &   $\sim$0.10$\pm$0.03 &   $\sim$1       &   $\sim$0.1      &   1.3$\pm$0.1   &   (23,25)   &   Centre \\
NGC~253           &   SB+Sy2+H~{\small II}   & $\sim$0.1      & $\sim$0.1-0.3       & $\sim$0.2-1.0 & $\sim$0.05-0.3 & 0.8$\pm$0.1   & (10,11)   & Centre \\
IRAS~210293       &   SB?          &   0.06$\pm$0.01  &   0.11$\pm$0.02       &      --	  &      --	       &   --            &   (20)      &   Centre \\
NGC~660           &   SB?+Sy2/L+H~{\small II} &   0.07$\pm$0.01  &   0.05$\pm$0.01      &   0.7$\pm$0.3   &   0.05$\pm$0.02  &   1.0$\pm$0.3   &   (20,21)   &   Centre \\  	   
NGC~891	          &   H~{\small II}          &   $\sim$0.2      &   --                  &   --            &   0.020$\pm$0.005&   1.1$\pm$0.5   &   (29,31)   &   Centre \\
NGC~986           &   SB+H~{\small II}       &   0.10$\pm$0.01  &   0.06$\pm$0.01       &   0.9$\pm$0.2   &   0.09$\pm$0.02  &   -             &   (20)      &   Centre \\   
NGC~1808          &   Sy2          &   0.06$\pm$0.01  &   0.07$\pm$0.01       &   1.0$\pm$0.2   &   0.06$\pm$0.01  &   0.56$\pm$0.04 &   (20,21)   &   Centre \\ 	  
NGC~2146          &   SB+H~{\small II}       &   0.08$\pm$0.01  &   0.10$\pm$0.01       &   0.8$\pm$0.2   &   0.06$\pm$0.01  &   0.77$\pm$0.05 &   (4,20,21) &   Centre \\	 
NGC~2369$^\dag$   &   SB?          &   --             &   --                  &   --            &   0.04$\pm$0.01  &   1.0$\pm$0.4   &   (29)      &   Centre \\
NGC~2903	  &   SB?+H~{\small II}      &   --             &   --                  &   --            &   0.020$\pm$0.005&   0.4$\pm$0.1   &   (29)      &   Centre \\
NGC~3256          &   SB+H~{\small II}       &   0.03$\pm$0.01  &   0.10$\pm$0.04       &   2.0$\pm$0.4   &   0.06$\pm$0.01  &   --            &   (20)      &   Centre \\ 	
NGC~4355	  &   Sy2          &   --             &   --                  &   --            &   0.8$\pm$0.2    &   0.08$\pm$0.01 &   (29)      &   Centre  \\
NGC~6946          &   SB+Sy2+H~{\small II}   & $\sim$0.05-0.1 & --                  & $\sim$1       & $\sim$0.2      & 1.2$\pm$0.1   & (3,4,8)   & Centre \\
NGC~7552          &   SB+L+H~{\small II}     &   0.07$\pm$0.01  &   0.11$\pm$0.02       &   1.1$\pm$0.2   &   0.08$\pm$0.01  &   0.9$\pm$0.2   &   (20,21)   &   Centre \\ 	
\hline                  
\end{tabular}
\end{center}
Notes: T is the numerical Hubble type. Sources marked with a
``$\star$'' show a particular gas chemistry with unusually high
HCN/$^{12}$CO, HCN/$^{13}$CO and/or HCN/HCO$^+$ ratios; sources marked
with a ``\dag'' or ``\dag\dag'' are LIRGs and ULIRGs
respectively. Values marked with a ``$\sim$'' or given as a range
indicate either values averaged over (or a range of values for)
several positions/observations or values for which an accurate
estimate of the uncertainty was not possible. $^a$ Sy=Seyfert;
L=LINER, SB= starburst, H~{\small II}=star formation bursts/H~{\small
II} regions, classifications taken from NED; $^b$ References: (1)
\citet*{wil00}, \citet{wild97} and \citet{wik97}; (2) this paper; (3)
\citet{kri08}; (4) Krips et al.\ (2010, in prep.); (5) \citet{pet03};
(6) \citet{isr09a}; (7) \citet*{isr06}; (8) \citet{pag01}; (9)
\citet{mat00}; (10) \citet{sak06}; (11) \citet{sor01}; (12)
\citet{wils97}; (13) \citet{tos07}; (14) \citet{bro05}; (15)
\citet{pet98}; (16) \citet{bol05}; (17) \citet{hei99}; (18)
\citet{chin97}; (19) \citet{gar08}; (20) \citet{aalto95}; (21)
\citet{baan08} (and references therein); (22) \citet{mei08}; (23)
\citet{ngrieu92}; (24) \citet{isr99}; (25) \citet{isr01}; (26)
\citet{isr09b}; (27) \citet{hen94}; (28) \citet{wan04}; (29)
\citet{baan08}; (30) \citet{glen01}; (31) \citet{saka97}; (32)
\citet{wil08}. $^c$ Regions in which the line ratios where determined:
Centre = central $\la10$~kpc of a galaxy; CND = circumnuclear disc,
i.e.\ a radius $<1$~kpc; GMC = average over several giant molecular
clouds; SA = spiral arms. $^d$ Ratio determined from absorption
lines. $^e$Ratios determined from interferometric observations.
\end{table*}

\begin{table*}
\contcaption{}
\begin{center}
\begin{tabular}{lcccccccc}     
\hline                  
Source & Activity Type$^a$ 
            & $\frac{^{13}{\rm CO(J=1-0)}}{^{12}{\rm CO(J=1-0)}}$  
            & $\frac{^{13}{\rm CO(J=2-1)}}{^{12}{\rm CO(J=2-1)}}$  
            & $\frac{{\rm HCN(J=1-0)}}{^{13}{\rm CO(J=1-0)}}$  
            & $\frac{{\rm HCN(J=1-0)}}{^{12}{\rm CO(J=1-0)}}$  
            & $\frac{{\rm HCN(J=1-0)}}{{\rm HCO^+(J=1-0)}}$  
          & Ref$^a$  & Region$^b$ \\
\hline    

\multicolumn{9}{c}{Starbursts: 0 $<$ T $<$ +8}\\[0.1cm]
  NGC~7771$^\dag$ &   SB+H~{\small II}       &   --             &   --                  &   --            &   $\sim$0.05     &   $\sim$1.0     &   (29)      &   Centre \\
  UGC~2855        &   ?            &   0.08$\pm$0.02  &   0.11$\pm$0.01       &       --        &       --         &   --            &   (20)      &   Centre \\
  IC~860$^\dag$          &   H~{\small II}          &   --      &   --         &   -- &   0.08$\pm$0.01 &   0.5$\pm$0.2   &   (29) &   Centre \\
  Arp~220$^{\dag\dag}$   &   SB+Sy2/L+H~{\small II} &   $<$0.05 &   $\sim$0.05 &   -- &   $\sim$0.08    &   0.46$\pm$0.09 &   (29) &   Centre \\
  IRAS~22025+4205$^\dag$ &   SB?          &   --      &   --         &   -- &   0.15$\pm$0.03 &   0.42$\pm$ 0.2 &   (29) &   Centre \\[0.15cm]

\multicolumn{9}{c}{SF-spirals: 0 $<$ T $<$ +8}   \\[0.1cm]
M33       &   H~{\small II} & $\sim$0.10    & $\sim$0.14 &  --           & --              & --          & (12)    & GMC in SA\\
M31       &   L?  & $\sim$0.13    & --         & $\sim$0.15    & $\sim$0.02      & $\sim$0.9   & (13,14) & GMC in SA\\
  Maffei~2  &   NONE  &   0.02-0.1      &   --         &   $\sim$0.4-1.3 &   $\sim$0.02-0.13 &   2.4$\pm$0.4 &   (22,23) &   GMC in SA\\[0.15cm]

\multicolumn{9}{c}{Dwarf galaxies: +8 $\leq$ T $\leq$ +10 }\\[0.1cm]
IC10 &   SB   & --         & $\sim$0.1     &  --           & --              & --            & (15)       & GMC \\
LMC  &   NONE & $\sim$0.1  & $\sim$0.2     & $\sim$0.2-0.4 & $\sim$0.03-0.06 & $\sim$0.3-0.7 & (16,17,18) & GMC \\
SMC  &   NONE & $\sim$0.1  & $\sim$0.1-0.2 & $\sim$0.3-0.4 & $\sim$0.02      & $\sim$0.4     & (16,17,18) & GMC \\ [0.15cm]

\multicolumn{9}{c}{Peculiar galaxies and/or mergers}\\[0.1cm]
  NGC~6240$^\dag$      &   Sy2/L  &   $\sim$0.02 &   $\sim$0.01 &   --     &   $\sim$0.09      &   $\sim$0.8   &   (29)    &   Centre \\	
  Mrk~273$^{\dag\dag}$ &   Sy2/L  &   --         &   --         &   --     &   $\sim$0.2       &   $\sim$ 0.4  &   (29)    &   Centre \\[0.2cm]

  NGC~3620             &   SB?    &   --         &   --         &   --     &   0.08$\pm$0.03   &   $<$0.4      &   (29)    &   Centre \\
  IC~1623$^\dag$       &   ?      &   --         &   --         &   --     &   0.040$\pm$0.006 &   1.2$\pm$0.2 &   (29)    &   Centre \\
  Arp~55$^\dag$        &   L+H~{\small II}  &   --         & $<$0.04        &   --     &   $\sim$0.1       &   $\sim$0.8   &   (29,32) &   Centre \\
  Arp~193$^\dag$       &   L+H~{\small II}  &   --         &   --         &   --     &   $\sim$0.03      &   $\sim$1.9   &   (29)    &   Centre \\
  Arp~299A$^\dag$      &   H~{\small II}    &   $\sim$0.04 &   $\sim$0.03 &  $\sim$1 &   0.040$\pm$0.006 &   1.6$\pm$0.3 &   (29)    &   Centre \\
  Arp~299B$^\dag$      &   H~{\small II}    &   $\sim$0.04 &   $\sim$0.03 &  $\sim$1 &   0.050$\pm$0.004 &   $\sim$0.8   &   (29)    &   Centre \\
  IRAS~12112+0305$^\dag$&   L+H~{\small II} &   --         &   --         &   --     &   $\sim$0.07      &   $\sim$0.6   &   (29)    &   Centre \\     
  IRAS~15107+0724      &   H~{\small II}    &   --         &   --         &   --     &   0.11$\pm$0.03   &   0.5$\pm$0.2 &   (29)    &   Centre \\ 		

\hline                  
\end{tabular}
\end{center}
\end{table*}

%
\section{OBSERVATIONS}
\label{sec:obs}
Using the IRAM 30m telescope at Pico Veleta, Spain, in August 2008 and
April 2010, we observed the $^{13}$CO(J=1--0), $^{13}$CO(J=2--1),
HCN(J=1--0) and HCO$^+$(J=1--0) lines in the centres of our four
target galaxies. In 2009, we used the AB set of receivers with the $1$
and $4$~MHz backends at $3$ and $1.4$~mm (89--110\,GHz and 22\,GHz),
respectively, while we used the Eight Mixer Receiver (EMIR) with the
Wideband Line Multiple Autocorrelator (WILMA) backend in 2010.  The
weather was exceptionally good in both years, with a
precipitable water vapour at $3$~mm of $\leq5$~mm during
$\approx90$~per cent of the time and $5$--$15$~mm during
$\approx10$~per cent of the time, corresponding to system temperatures
$T_{\rm sys}=100$--$200$~K at $3$~mm and $\approx300$~K at
$1.4$~mm. We discarded all $1.4$~mm data obtained at $3$~mm
precipitable water vapour of $5$--$15$~mm while we kept the
corresponding 3~mm data as the system temperatures were still in an
acceptable range (i.e., Tsys$\simeq$250~K). We spent $2$--$5$~hr on
source per target and frequency, resulting in an rms noise level (in
T$_{\rm mb}$ scale) of $\approx0.5$--$1$~mK for HCN and HCO$^+$
(binned to $\approx30$~km~s$^{-1}$) and of $\approx2$--$3$~mK for
$^{13}$CO (binned to $\approx20$~km~s$^{-1}$). Pointing and focus
tests were carried out every $1.5$--$2$~hr on a nearby planet
($<30^\circ$; Mars and/or Saturn) or a nearby strong quasar
($<10^\circ$; 0924+392). The pointings were consistent with each other
within $\approx1$--$3$~arcsec throughout the observations. This is
entirely sufficient given the beam sizes (half power beam width, HPBW)
of $27$~arcsec for HCN(J=1--0) and HCO$^+$(J=1--0), $22$~arcsec for
$^{13}$CO(J=1--0) and $11$~arcsec for $^{13}$CO(J=2--1).

To remove the continuum from each spectrum, a baseline of polynomial
order $0$ or $1$ was fit to the line-free regions of each scan and
subtracted, and the scans were thereafter averaged.  We measured the
peak intensities, central velocities, full width half maxima (FWHM)
and velocity-integrated fluxes of the detected lines by fitting
Gaussian profiles to the data. Some of the lines clearly exhibit
multiple peaks, so we compared line intensities obtained from both
multiple and single Gaussian fits. As the (summed) line intensities
agreed well with each other, we will use the values from the single
Gaussian fits in the following.

Antenna temperatures ($T_{\rm a}^\star$) have been converted to main
beam temperatures ($T_{\rm mb}$) by dividing the antenna temperatures
by $\eta\equiv B_{\rm eff}/F_{\rm eff}$, where $B_{\rm eff}$ and
$F_{\rm eff}$ are the beam and forward efficiencies, respectively
(i.e., $T_{\rm mb}=T_{\rm a}^\star/\eta$; $\eta=0.78/0.95=0.82$ for
HCN and HCO$^+$, $\eta=0.75/0.95=0.79$ for $^{13}$CO(J=1--0) and
$\eta=0.55/0.91=0.60$ for $^{13}$CO(J=2--1)\,). To convert the
measured brightness temperatures ($T_{\rm mb}$[K]) to fluxes (S[Jy])
via S[Jy]$=f_{\rm conv}\,T_{\rm mb}$[K], the following conversion
factors must be used: $f_{\rm conv}=4.9$~Jy~K$^{-1}$ for HCN(J=1--0)
and HCO$^+$(J=1--0), $f_{\rm conv}=5.0$~Jy~K$^{-1}$ for
$^{13}$CO(J=1--0) and $f_{\rm conv}=5.8$~Jy~K$^{-1}$ for
$^{13}$CO(J=2--1).

%
%
\begin{table*}
\caption{Properties of the molecular lines detected.}
\label{tab:tab4}      
\centering          
\begin{tabular}{l c c c c c c c c}     
  \hline
  Source  &  $L_{\rm IR}$     & $L_{\rm HCN(1-0)}$    & $L_{\rm ^{13}CO(1-0)}$ & $L_{\rm ^{12}CO(1-0)}$  & $M_{\rm dense}$(H$_2$)
  & $M$(H$_2$)  & $L_{\rm IR}/M_{\rm dense}$(H$_2$) & $L_{\rm IR}/M$(H$_2$) \\
  & ($10^9$~$L_\odot$) & ($10^8$~K~km~s$^{-1}$~pc$^2$) & ($10^8$~K~km~s$^{-1}$~pc$^2$) & ($10^8$~K~km~s$^{-1}$~pc$^2$)
  & ($10^8$~$M_\odot$) & ($10^8$~$M_\odot$) & ($L_\odot/M_\odot$) & ($L_\odot/M_\odot$)  \\
  \hline 
  NGC~3032 &  3.9 & 0.03    & 0.08 &  0.8 & 0.3   & 3.8 & 130 & 10 \\
  NGC~4150 &  0.7 & $<$0.008 & 0.02 &  0.2 & $<$0.08  & 1.0 & $>$90 &  7 \\
  NGC~4459 &  2.2 & 0.04    & 0.17 &  0.6 & 0.4   & 3.0 &  55 &  7 \\
  NGC~4526 &  5.8 & 0.08    & 0.33 &  1.2 & 0.8 & 5.7 &  72 & 10 \\
  \hline
\end{tabular}
Notes: $L$ = luminosity, $M$ = mass. The infrared luminosity (L$_{\rm
  IR}$) is determined from the $12$, $25$, $60$ and $100$\,$\mu$m
fluxes using L$_{\rm IR}$~=~4$\pi$D$_{\rm L}$$^2$~F$_{\rm
  IR}$[L$_\odot$], with F$_{\rm
  IR}$[W~m$^{-2}$]~=~1.8$\times$10$^{-14}$~(13.48~f$_{\rm 14 \mu
  m}$~+~5.16~f$_{\rm 25 \mu m}$~+~2.68~f$_{\rm 60 \mu m}$~+~f$_{\rm
  100 \mu m}$) and D$_{\rm L}$= luminosity distance. The infrared
luminosities were derived from {\it IRAS} fluxes that were taken from
the NED. The mass of the dense molecular gas is derived using $M_{\rm
  dense}$(H$_2$)$~=~10~L_{\rm
  HCN}$~$M_\odot$~(K~km~s$^{-1}$~pc$^{2}$)$^{-1}$ \citep*{rad91},
while the entire molecular gas mass is derived using
$M$(H$_2$)$=4.8\times L_{\rm
  ^{12}CO}$~$M_\odot$~(K~km~s$^{-1}$~pc$^{2}$)$^{-1}$ \citep[][and
  references therein]{you91}.
\end{table*}

%
%
\begin{figure*}
\begin{center}
\resizebox{\hsize}{!}{\rotatebox{-90}{\includegraphics{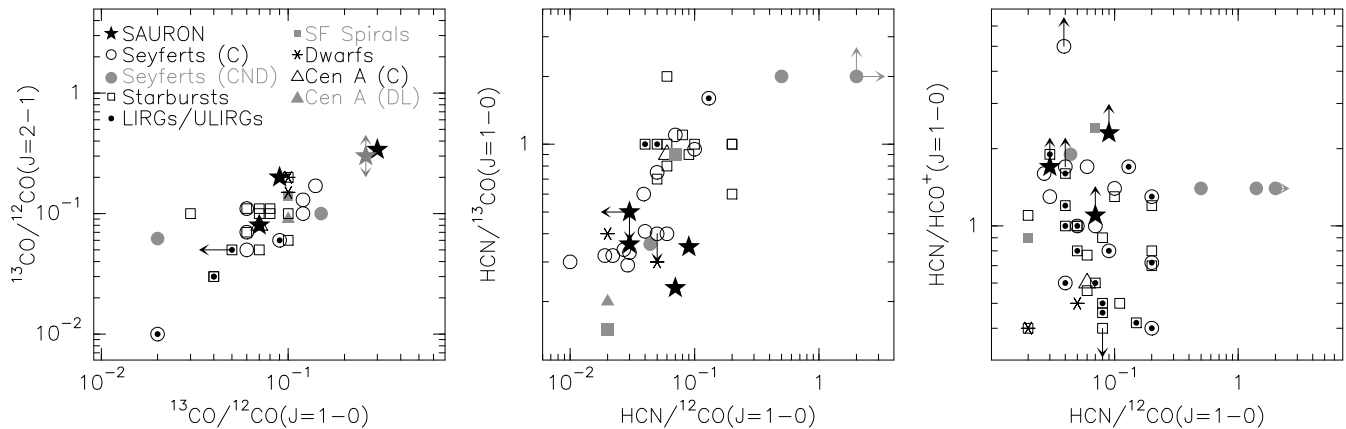}}}
\caption{Molecular line ratios of the {\tt SAURON} early-type galaxies
  compared to those of galaxies of different Hubble and activity types
  (see Table~\ref{tab:tab3}). Errors are of the order of the symbol
  sizes for most sources. For the labels: C = centre; CND =
  circumnuclear disc; DL = dust lane. LIRGs/ULIRGs are additionally
  highlighted by small filled circles. Given the uncertainty of the
  $^{13}$CO/$^{12}$CO(J=2--1) line ratio for NGC~4526, its
  corresponding symbol is plotted in grey and marked additionally with
  errors pointing up and down.}
\label{fig:fig2}
\end{center}
\end{figure*}

\section{RESULTS AND DISCUSSION}
\label{sec:results}

We detected $^{13}$CO(J=1--0) and $^{13}$CO(J=2--1) emission in four
out of four sources and HCN(J=1--0) emission in three out of four
sources, while HCO$^+$ emission is undetected in all four sources. The
spectra of the individual lines for each galaxy are plotted in
Fig.~\ref{fig:fig1}. We also overlay the $^{12}$CO emission from
\citet{com07} on our $^{13}$CO detections for a direct comparison. The
line parameters obtained from Gaussian fits to the data are listed in
Table~\ref{tab:tab2} and compared to the $^{12}$CO data from
\citet{com07}. All the line profiles appear to agree well with each
other for each individual galaxy. Only the $^{13}$CO(J=2--1) line in
NGC4459 and NGC4526 shows some asymmetry, with a stronger blueshifted
peak, but this can probably be attributed to the smaller beam at
$1.4$~mm and thus a different beam filling at 3 and 1.4~mm. While the
$^{13}$CO line emission in all galaxies except NGC~4526 strongly
resembles that of $^{12}$CO, the $^{13}$CO(J=2--1) line in NGC~4526
differs from that of $^{12}$CO(J=2--1). This is most likely due to a
slightly different pointing during the two observations, which will be
considered for the line ratios in the following discussion. It also
indicates that the emission in NGC~4526 (and probably NGC~4459) is
slightly larger than the beam size at 1~mm. The fitted heliocentric
velocities also agree with each other for all galaxies except
NGC~4526, where the measured velocities differ from each other by
$\approx$90~km~s$^{-1}$. However, this difference can be explained by
the Gaussian fit to the asymmetric line profile of the $^{12}$CO
emission done in \citet{com07}, which favours the stronger redshifted
peak and thus shifts the line centre towards higher velocities (see
also Fig.~\ref{fig:fig1}). If one calculates the central velocities
from the full widths at zero maximum (FWZM), one finds a more
consistent value of $\approx620$~km~s$^{-1}$.

%
%
%
\begin{figure*}
\begin{center}
\resizebox{\hsize}{!}{\rotatebox{-90}{\includegraphics{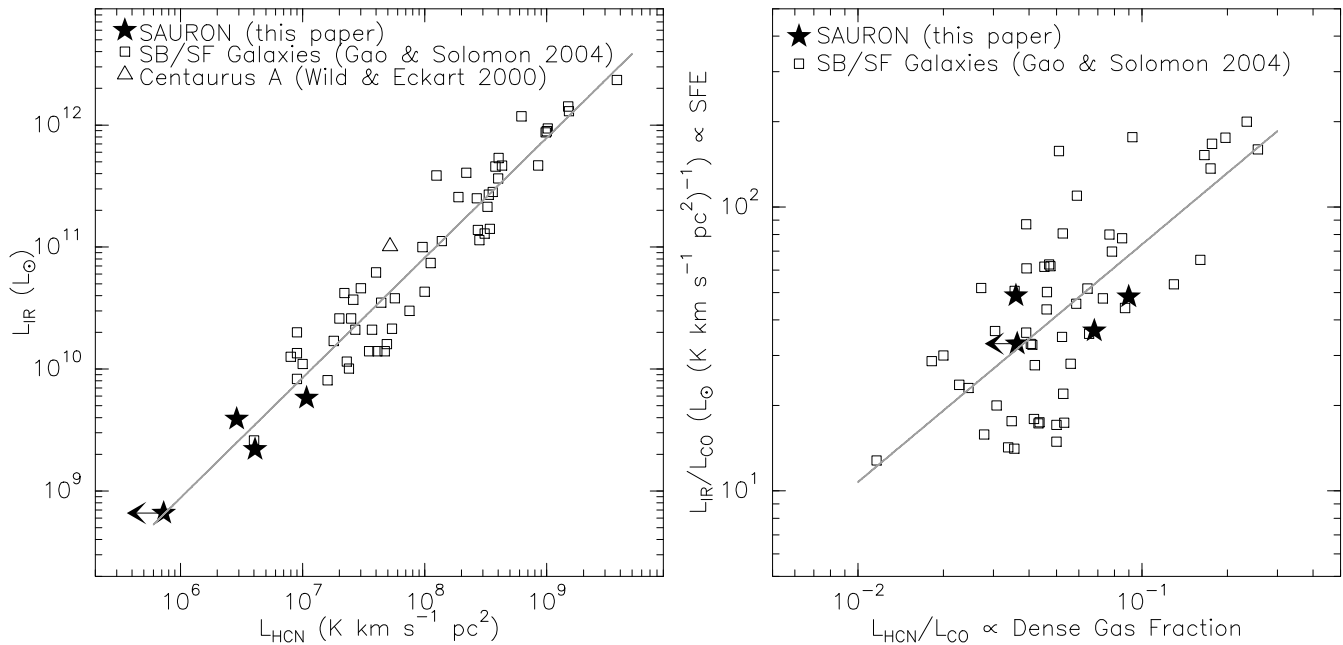}}}
\caption{Infrared and HCN luminosities. {\it Left:} Infrared
  luminosity as a function of the HCN(J=1-0) luminosity. {\it Right:}
  Dense gas fraction versus star formation efficiency.}
\label{fig:fig3}
\end{center}
\end{figure*}

We calculated line ratios between the total of $6$ molecular lines
observed for each galaxy (see Table~\ref{tab:tab3} and
Fig.~\ref{fig:fig2}). To avoid systematic errors due to beam dilution
effects, we only determined line ratios at similar frequencies, i.e.\
observed with similar beam sizes. Please note that we consider the
line ratio between the $^{13}$CO(J=2--1) and $^{12}$CO(J=2--1) line
emission as uncertain given their different line profiles (see
discussion in the previous paragraph). The line ratios are compared to
those in galaxies of different Hubble types in Table~\ref{tab:tab3}
and Figure~\ref{fig:fig2}. The list of galaxies of different Hubble
types should be regarded as representative of each type and does not
attempt to be complete. We explicitly state the region of each galaxy
where the line ratios have been taken. This is especially important
for the very nearby galaxies such as M33, M31 and the dwarf galaxies,
for which the observations already resolve giant molecular clouds
(GMCs) and cannot be regarded as averaging over the central regions of
each galaxy. In addition, previous studies indicate that molecular
line ratios can vary significantly within a galaxy
\citep*[e.g.,][]{wils97,hel97,bro05,kri07}, especially for galaxies
with an active galactic nucleus (AGN). NGC~1068 and NGC~6951 are
prominent examples of completely different chemistry and excitation
conditions of the molecular gas close to the AGN and in the (star
formation-dominated) large-scale spiral arms
\citep*[e.g.,][]{ster94,kri07,kri08}.

Although the $^{13}$CO/$^{12}$CO line ratios appear to be quite
similar among galaxies of different Hubble types, with values in
general around $0.1$-$0.2$, certain trends can nevertheless be
identified. While the starburst galaxies and (U)LIRGs seem to populate
the lower end of the line ratio distribution (with values
$\lesssim$0.1), three of the four SAURON galaxies lie at the higher
end of it, with values larger than $0.2$ in at least one of the two
transitions (see Fig.~\ref{fig:fig2}). \citet{cas92} suggest that the
low $^{13}$CO/$^{12}$CO ratios in starburst galaxies are not a
consequence of excitation conditions and/or optical depth effects, but
are rather due to an actual change in the $^{13}$CO/$^{12}$CO
abundance ratio itself, either by overabundant $^{12}$CO or
underabundant $^{13}$CO. As numerous new massive stars are formed
during a starburst, the interstellar medium will be preferentially
enriched by nucleosynthesis of $^{12}$C, whereas $^{13}$C remains
unaffected. Also, at least in the case of a merger-induced starburst,
the infall of large-scale gas to the centre will result in a dilution
of the central $^{13}$CO gas, as the infalling gas is poor in
$^{13}$CO. As for the high $^{13}$CO/$^{12}$CO line ratios in three of
the four {\tt SAURON} galaxies, optical depth may well play a
significant role, with e.g., optically thick $^{12}$CO emission.

The HCN/$^{13}$CO and HCN/$^{12}$CO ratios are similar to those of the
(star formation-dominated) central regions of nearby Seyfert and
spiral galaxies as well as those of GMCs in dwarf galaxies (see
Fig.~\ref{fig:fig2}), but they are quite different from (i.e., smaller
than those of) nearby starburst galaxies and the (AGN-dominated)
circumnuclear discs (CNDs) of some Seyfert galaxies (CNDs being
smaller than the central regions considered in the Seyferts).

The {\tt SAURON} early-type galaxies stand out most when considering
the HCN/HCO$^+$ ratio, varying most significantly again not as a
function of Hubble type but rather of activity type
\citep[Fig.~\ref{fig:fig2}; see also][]{kri08}. Here, the {\tt SAURON}
galaxies strongly resemble the (AGN-dominated) circumnuclear regions
of some Seyfert (e.g.\ NGC~1068, NGC~6951 and M51) and some
(ultra-)luminous infrared galaxies ((U)LIRGs). However, they differ
from most starburst and the dwarf galaxies (at least the Magellanic
Clouds) as well as Centaurus~A (Cen~A). Nevertheless, because of the
high HCN/$^{13}$CO and HCN/$^{12}$CO ratios of these Seyfert galaxies,
the high HCN/HCO$^+$ ratios in the {\tt SAURON} early-type galaxies
probably have a different origin than that in the Seyferts. While the
high HCN/HCO$^+$ ratios in the Seyfert galaxies are believed to be a
consequence of enhanced HCN emission
\citep[e.g.,][]{ster94,use04,kri07,kri08}, those in the {\tt SAURON}
galaxies appear to be due to lowered HCO$^+$ emission instead.

HCO$^+$ is known to be enhanced over HCN in galaxies with strong
starburst activity, with a likely dependence on the age of the star
formation burst, i.e.\ the older the star formation burst the lower
the HCN/HCO$^+$ ratio (see the examples of the evolved starburst
galaxy M82 and the young starburst NGC~6946; \citealt{kri08} and
references therein). This effect is thought to be a consequence of the
increased rate of supernova explosions and/or the increased importance
of photo-dissociation regions (PDRs) in more evolved starburst
galaxies. The high HCN/HCO$+$ ratios observed here in the four {\tt
  SAURON} targets may thus indicate that supernova explosions (or,
equally, PDRs) only play a minor role in the gas chemistry of these
early-types.

It is interesting to note that the molecular line ratios of most of
the {\tt SAURON} early-type galaxies are not very similar to those of
the elliptical galaxy Cen~A. Cen~A rather seems to favour line ratios
close to those of the dwarf galaxies in our list. This is not entirely
surprising however as the molecular gas in Cen~A, mostly located in
the minor-axis dust lane, has almost certainly been captured during a
merger event with a gas-rich dwarf or small disc galaxy
\citep*[e.g.,][]{mal83}.

When comparing the HCN luminosity to the IR luminosity (a measure of
star formation; see Fig.~\ref{fig:fig3}), one finds that the four {\tt
  SAURON} sources nicely follow the IR--HCN correlation of
star-forming galaxies, firmly extending the relation to lower
luminosities. Plotting both of these luminosities (IR and HCN)
normalized by the $^{12}$CO luminosity (thus measuring star formation
efficiency against dense gas fraction), our early-type galaxies again
lie comfortably on the relation \citep[e.g.,][]{sol92,gao04a,gao04b},
having both relatively average dense gas fractions and star formation
efficiencies (SFEs). This suggests that most of the star formation in
these four SAURON early-type galaxies is taking place in dense gas
regions, as in other nearby (spiral) galaxies. The dense gas fraction
is however lower than what is found in truly starbursting galaxies.

\section{SUMMARY AND CONCLUSIONS}
\label{sec:conclusions}
We detected significant $^{13}$CO in four out of four and HCN emission
in three out of four {\tt SAURON} early-type galaxies, while no
HCO$^+$ emission was found in any of the four sources. We find some
pronounced differences in the line ratios of the {\tt SAURON} galaxies
when compared to other nearby galaxies of different Hubble and
activity types. In particular, $^{13}$CO/$^{12}$CO appears slightly
enhanced in three {\tt SAURON} galaxies compared to other galaxy
types. This may indicate different molecular gas excitation conditions
and/or chemistry in these sources.  The closest resemblance to our
four nearby {\tt SAURON} galaxies is found in nearby star-forming
galaxies, including Seyferts and dwarfs. This and the pronounced
differences with starburst galaxies suggest that the four {\tt SAURON}
galaxies exhibit star-formation rates and efficiencies more similar to
quiescent, normal star-forming galaxies than to starburst galaxies,
and that they do not reach the high dense molecular gas fraction found
in starbursts. Also, according to the high ($>$1) HCN/HCO$^+$ ratios,
PDRs and/or supernovae explosions do not seem to play an important
role in the chemistry of the molecular gas in our four targets.

The three {\tt SAURON} galaxies observed in HCN nicely
follow the same physical laws concerning star formation and dense
molecular gas as other galaxies, at least when infrared radiation is
used as the star formation tracer.

Although the four {\tt SAURON} galaxies studied in this paper form a
nice subset of the {\tt SAURON} sample with respect to their general
characteristics, they are not equally representative with respect to
their molecular gas properties.  They have been chosen for their
strong $^{12}$CO emission and could thus in principal be extreme cases
within the sample. Follow-up observations of the $^{13}$CO, HCN and
HCO$^+$ emission in a much larger sample of nearby early-type galaxies
are currently underway based on the success of this pilot study. These
will eventually allow us to confirm (or discard) the trends found in
this pilot project and put them on a sounder statistical basis.

%
%
\section*{Acknowledgments}
We thank the IRAM staff at the IRAM 30~m telescope at Pico Veleta for
their help during the observations. We would like to thank various
members of the {\tt SAURON} team for useful discussions throughout
this project and the referee for a thorough review of the paper.  We
acknowledge the usage of the HyperLeda database
\citep[][http://leda.univ-lyon1.fr]{pat03}.  This research has made
use of the NASA/IPAC Extragalactic Database (NED) which is operated by
the Jet Propulsion Laboratory, California Institute of Technology,
under contract with the National Aeronautics and Space Administration.
%
%

%
\end{document}